# Structural, electrical, magnetic, and flux pinning properties of YBCO/Ni superconducting composites: analyses and possible explanations

Ahmed M. Abd El-Aziz[a], Hesham A. Afifi[a], Ibrahim Z. Hager[b], Nadia S. Abdel Aal[a] and S. H. Naqib[c*]

[a] Ultrasonics Lab, National Institute for Standards, Giza, Egypt
[b] Department of Physics, Faculty of Science, Menoufia University, Shebin El-Koum, Egypt
[c] Department of Physics, University of Rajshahi, Rajshahi 6205, Bangladesh
[*] Corresponding author salehnaqib@yahoo.com

## ABSTRACT

Samples of superconducting composite $(100-x)YBa_2Cu_3O_{7-\delta} + xNi$ (x = 1, 2.5, 5, 7.5, 10 and 15 wt.%) were synthesized. Scanning electron microscope (SEM) images revealed that the grain size of YBCO reduces gradually as Ni content is increased. At the same time segregation of Ni as NiO in patches forms throughout the composite. At low Ni wt.%, it was found that the superconducting state properties improved and then dropped with higher Ni wt.%. Segregated patches of NiO on different length scales were observed. The change in the structural and superconducting state parameters of YBCO and the variation of the magnetic critical current density with improved flux pinning in some composites with Ni additives were ascribed to the Jahn-Teller (JT) distortions and pinning of vortices due to antiferromagnetic NiO patches within the samples.

## 1. Introduction

$YBa_2Cu_3O_{7-\delta}$ (YBCO) is among the most widely studied hole doped superconducting (SC) cuprates. This double $CuO_2$ layer compound is relatively easy to synthesize, the in-plane hole content can be varied by changing the oxygen level in the $CuO_{1-\delta}$ chains, the superfluid density of the fully oxygenated YBCO is quite high, and consequently this compound has very high critical current density ($J_c$) and high irreversibility magnetic field ($H_{irr}$)[1–8]. As far as potential applications are concerned, YBCO is one of the most promising hole doped high-$T_c$ cuprate. Because of increasing needs of ever higher values of $J_c$, $H_{irr}$, and improved mechanical properties for industrial applications, YBCO composites are made from different additives for superior electrical and magnetic properties[9–11]. The structure of the superconducting composites is characterized by inhomogeneous micro-defects that can act as artificial flux-pinning sites[9–11]. The effective pinning sites introduced into superconducting YBCO enhance $J_C$ in applied magnetic field that is useful for various high-field applications[9–13]. It has been found that nano-scale defects within the superconducting matrix can act as effective pinning centers for magnetic vortices in YBCO[1,3]. Flux pinning by inclusion of magnetic materials was proposed to strengthen pinning through the magnetic interaction between the defect and the flux lines[14–16]. However, transition elements like Fe, Ni and Co[15–23] were found to have a detrimental effect on the superconducting behavior of YBCO matrix due to significant cross-contamination of the pristine lattice. These impurities partially substitute the in-plane Cu-sites and therefore, affect the electronic ground state of YBCO at an undesirable level. The effect of nonmagnetic ions, such as Zn and Mg, were also extensively studied[15–18,23–26]. All these impurities, at least partly, substituted in the $CuO_2$ plane suppress the superconductivity in cuprates to different extents[27,28]. The decrease of $T_C$ by Ni doping for YBCO is comparatively smaller among these dopants[29,30]. In the present work, attempts have been made to include agglomerations of ferromagnetic metal Ni, which changes into oxide form NiO, in YBCO matrix. Different weight percentage of Ni has been used. The structural, morphological, charge transport, magnetic, and critical current density properties have been studied in detail. Unlike in most of the previous studies[1–3], where the focus has been placed on critical current density and flux pinning properties, we have tried to model the changes in the physical properties due to NiO additives in terms of Jahn-Teller (JT) distortion in this paper. The effect of inclusion of NiO on structural and electrical transport properties is interesting and worthy of study.

The organization of the rest of the paper is as follows. Experimental samples and techniques are described briefly in Section 2. Experimental results are presented and analyzed in Section 3. The results are discussed and a physical model has been introduced in Section 4. Finally, important conclusions of this study are summarized in Section 5.

## 2. Experimental Samples and Experimental Methods

$YBa_2Cu_3O_{7-\delta}$ samples were prepared from 99.9%-purity powders of $Y_2O_3$, $BaCO_3$ and CuO. A stoichiometric amount of the cationic ratio of Y:Ba:Cu = 1:2:3, respectively, was stirred in distilled water and then well mixed for 12 hrs. using a high power sonifier. The solution was dried at 90C until well mixed dry powders were obtained. The powders were then ground for 2 hrs. and sintered at 900C for 20 hrs. and then annealed at 550C for 15 hrs. in oxygen-enriched environment. The last process was repeated. A series of polycrystalline composite samples of $(100-x)YBa_2Cu_3O_{7-\delta} + xNi$ (with x = 0, 1, 2.5, 5, 7.5, 10 and 15wt.%) (henceforth, YBCO + xNi) were ground and then pressed into pellets (Ni metal powders with comparatively large grain size was used). The pellets of the composites were sintered at 900C for 20 hrs. and cooled to 550C where they were kept for 15hrs. at an oxygen-enriched environment. The X-ray diffraction (XRD) patterns of the samples were obtained via X-ray Diffractometer (PANalytical) with Cu $K_\alpha$ radiation ($\lambda$ = 1.54 Å). The XRD data were collected over the diffraction angle range $2\theta = 10°-80°$. Information regarding the microstructure was collected using a scanning electron microscope (SEM) (Quanta 250 FEG) with an Energy Dispersive X-ray (EDX) spectroscopy analyzer with an accelerating voltage of 30 kV. Measurements of temperature dependent resistivity, $\rho(T)$, were made employing the standard four-probe technique with a nano-voltmeter (Keithley 182) having $10^{-9}$V resolution. Voltage drop in the samples was measured with the current flow of 20 mA from a constant current source (Keithley 224). The temperature of the samples was varied by a cooling system with a temperature resolution of ± 0.1 K.

The magnetization vs. applied magnetic field (M-H curve) data of the composites was taken at ~ 79 K via a vibrating sample magnetometer (Lakeshore 7410 VSM). DC magnetic field was applied up to $20\times10^3$ Gauss. The field accuracy in gauss was 1% of the applied field. The magnetic moment measurement range was from $0.1 \times 10^{-6}$ emu to 1000 emu. The system has absolute accuracy better than 1% of the reading and reproducibility better than ±1%. The noise floor/sensitivity has been 0.1 μemu at 16.2 mm operating air gap with standard configuration.

## 3. Experimental Results and Analyses

### 3.1. XRD analysis

Figure 1(a) shows the XRD patterns of the samples YBCO + xNi (x = 0, 1, 2.5, 5, 7.5, 10 and 15 wt.%). The results show a predominant phase of YBCO with orthorhombic $P_{mmm}$ symmetry. Monoclinic phase of NiO is clearly observed with growing peaks as Ni content increases. Figure 1(b) shows the relative intensity at (001), (200) and (111) planes of NiO monoclinic phase with Ni wt.%.

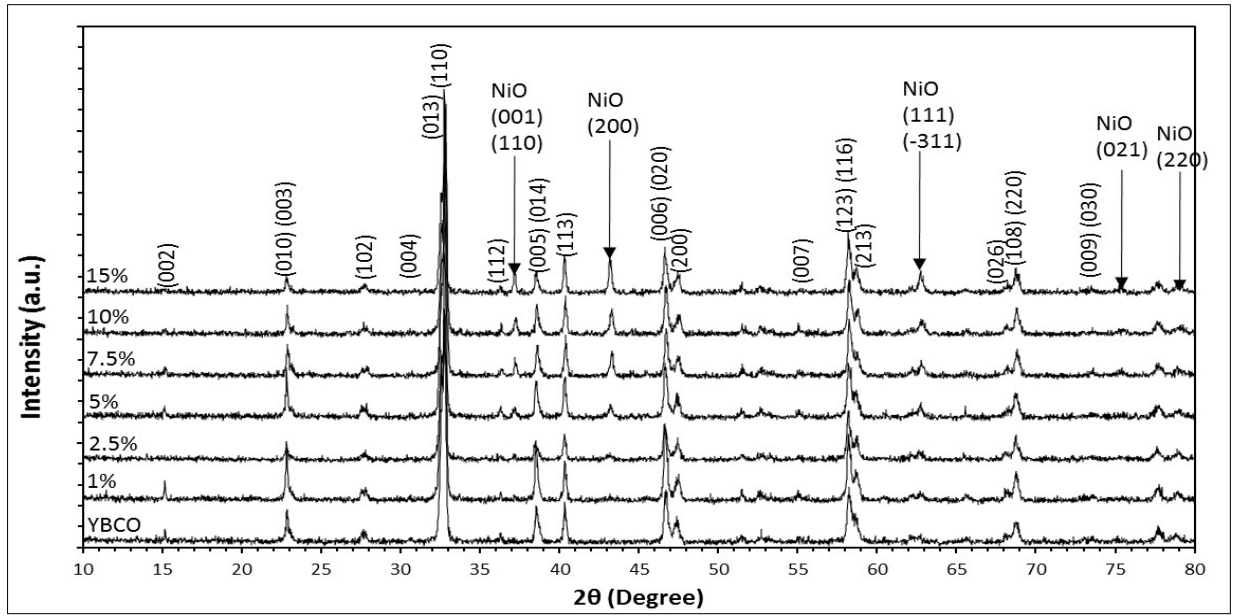

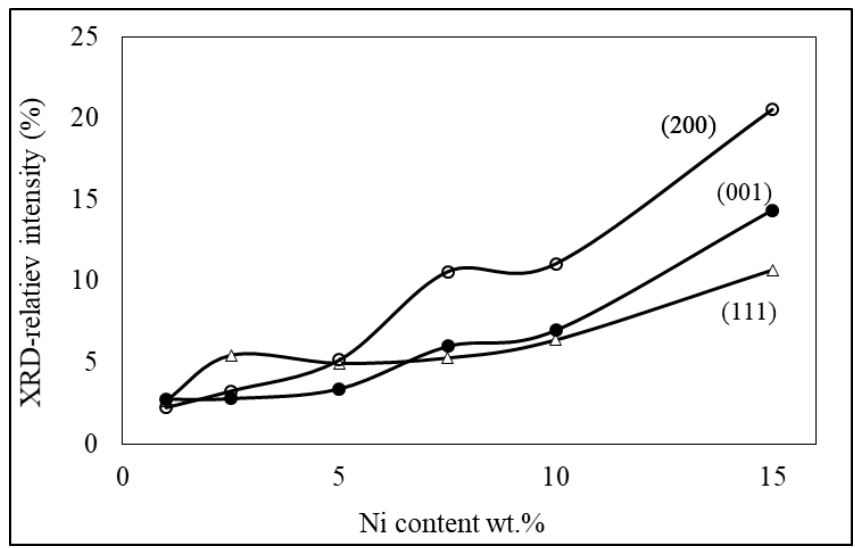

**Figure 1:** (a) XRD Patterns of YBCO + xNi samples (x = 0, 1, 2.5, 5, 7.5, 10, 15 wt.%) (b) XRD relative intensity from (200), (001) and (111) planes of NiO monoclinic phase vs. Ni wt.%.

It is clearly seen that spectral weights under the relevant diffraction peaks increase with increasing Ni content. From the analysis of the XRD data it can be stated that large quantities of Ni were incorporated, or agglomerated to be precise, as NiO in the YBCO. Table 1 exhibits the calculated lattice parameters a, b and c using the least square fits using the d values and (*hkl*) planes for orthorhombic unit cell structure.

**Table 1:** The lattice constants of YBCO unit cell and the c/a ratio with different Ni content.

| Ni (wt.%) | a (Å) | b (Å) | c (Å) | c/a |
|---|---|---|---|---|
| 0 | 3.830 | 3.890 | 11.680 | 3.050 |
| 1 | 3.826 | 3.882 | 11.682 | 3.053 |
| 2.5 | 3.832 | 3.903 | 11.690 | 3.050 |
| 5 | 3.830 | 3.893 | 11.669 | 3.047 |
| 7.5 | 3.834 | 3.877 | 11.640 | 3.036 |
| 10 | 3.820 | 3.886 | 11.664 | 3.054 |
| 15 | 3.826 | 3.885 | 11.655 | 3.046 |

From the XRD data, the orthorhombic peaks of YBCO in all composite samples were noted to have slightly different lattice constants. In other words, YBCO crystallites with slightly different lattice parameters (possibly due to slightly different oxygen content and/or due to surrounding NiO induced lattice strain and partial Ni substitution in the Cu sites) were formed within the same sample by Ni addition, i.e. different orthorhombic phases of different unit cell volumes (Å$^3$) and different c/a distortions, as seen in Figure 2. It is therefore, reasonable to assume that Ni ions were added at different levels in different places in the YBCO matrix in each of the samples. It should be noted that in an undistorted lattice with homogeneous composition, equivalent Miller indices (*hkl*) give same lattice constants. In the presence of local distortion this is not the case. Incorporating the SEM, EDX and resistivity results (presented in the next sections), it can be extrapolated that some of the Ni ions moved away from NiO patches in three-dimensional percolation process through YBCO crystal lattice substituting for the Cu-sites. It has been reported earlier that the $Ni^{2+}$ substituting $Cu^{2+}$ in Cu-O plane, expressed as $YBa_2(Cu_{1-y}Ni_y)_3O_{7-\delta}$, leads to a decrease in the bond distance between the apical O and the plane $Cu^{19,31–33}$. YBCO traces which are sticking to the NiO patches can be thought of as regions of high $Ni^{2+}$-doping and of diminishing c/a ratios. Whilst, the YBCO crystallites separated from NiO patches are thought of as regions with lower $Ni^{2+}$-doping where the c/a is comparatively larger. Rising number of $Ni^{2+}$ ions in Cu-plane sites, as a result of increasing Ni wt.% from 2.5 up to 7.5%, gives rise to steady decrement in c/a ratios. On the other hand, more Ni wt.%, namely 10 and then 15 %, gives rise again to increased c/a ratios relative to 7.5 % sample, as seen in Figure 2. It is important to note that c/a ratio has been employed to characterize the Jahn-Teller distortion of the oxygen pyramid around $Cu^{2+}$ (or $CuO_5$ pyramid)$^{34–36}$.

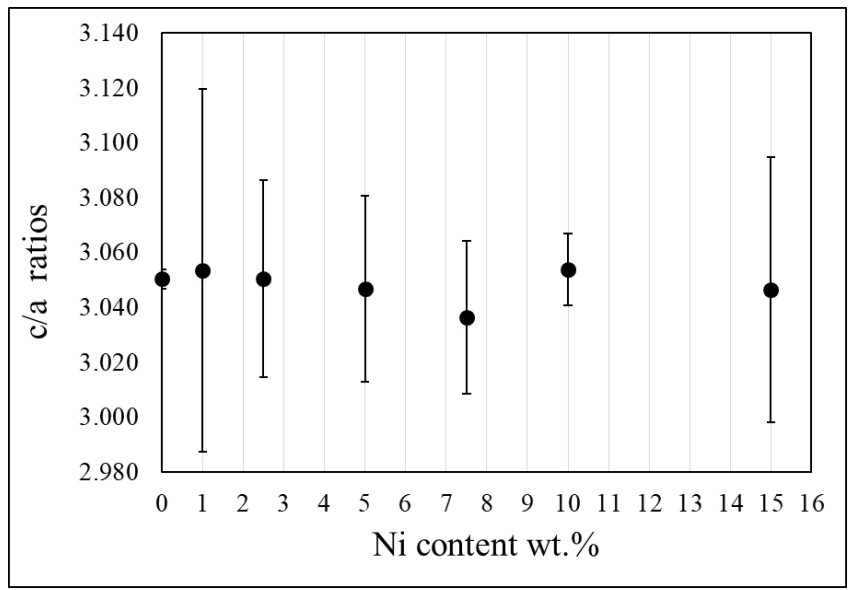

**Figure 2:** c/a ratios of YBCO orthorhombic phases as a function of Ni. wt.%.

## 3.2. Microstructural and EDX analysis

Figure 3 shows the SEM images of the YBCO + xNi (x = 0, 2.5, 5, 10 and 15 wt. %) composites, in addition to the EDX analysis for the positions marked in these images. From the SEM images, two distinct features are observed; the first one is that all the samples exhibit randomly oriented grains (with varied sizes) in all directions with the presence of some pores in between. Secondly, the segregation of the well-formed NiO in patches. EDX analysis for the positions X1 to X3 shows that all the compositional elements are in proper stoichiometric ratio. These data show straightforwardly how the NiO patches might affect the level of Ni incorporation into the YBCO matrix and offers possible explanation to the data presented in Table 1. Figure 3(X1) give the EDX analysis of position X1 shown in Figure 3(a) which reveals the elemental composition of pure YBCO. Figures 3(X2) and 3(X3) illustrate the EDX analysis for positions X2 and X3 (shown in Figure 3(b)). This verifies the high concentration of Ni in the *blackish* patch and the reduction in Ni concentration outside that patch. In other words, Ni is segregated into NiO patches and its concentration decreases as one moves away from those patches as indicated by Figure 3(X4). The linear EDX analysis of it is shown in Figure 3(d).

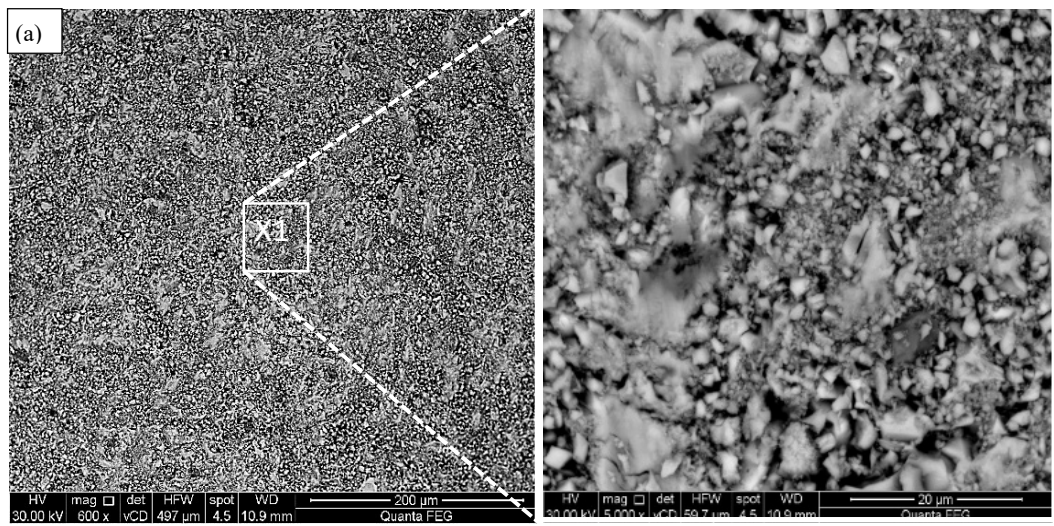

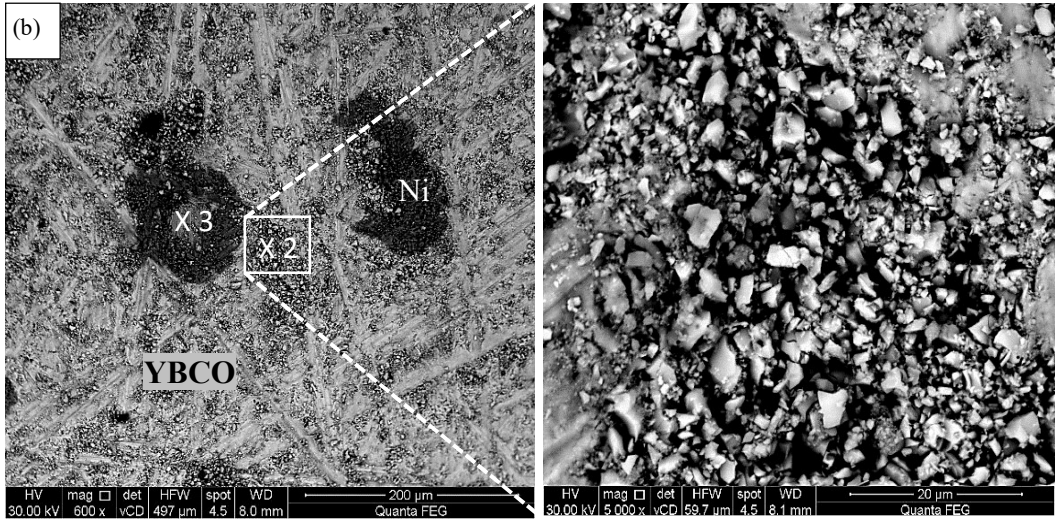

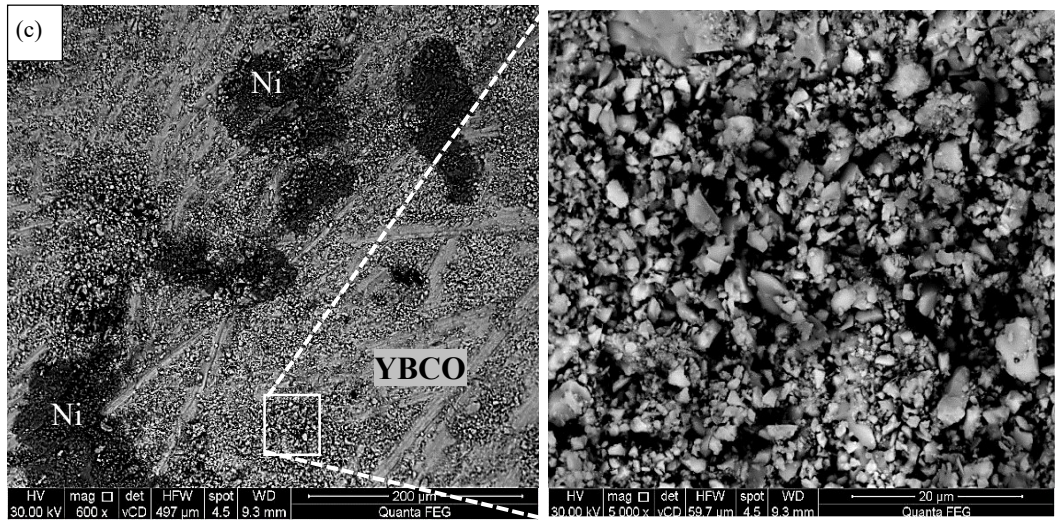

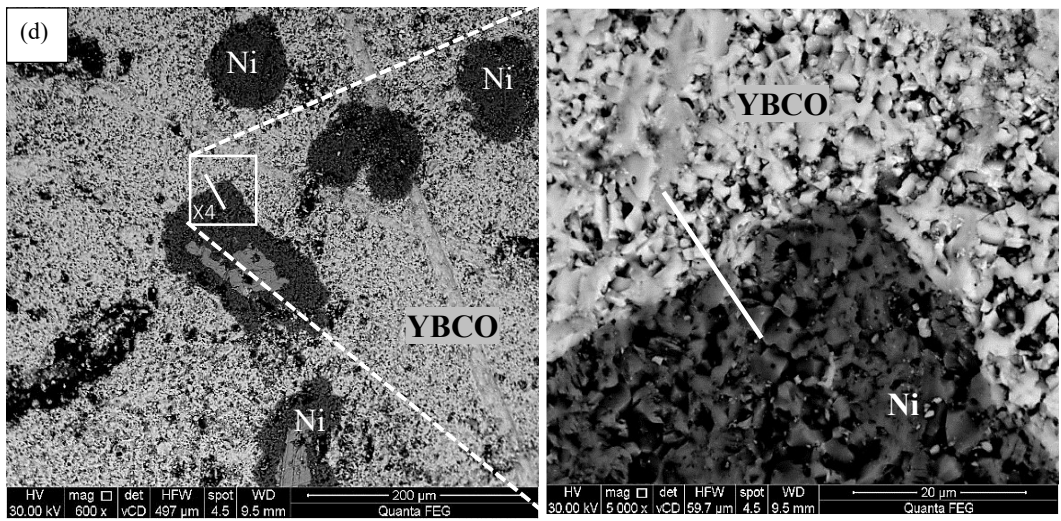

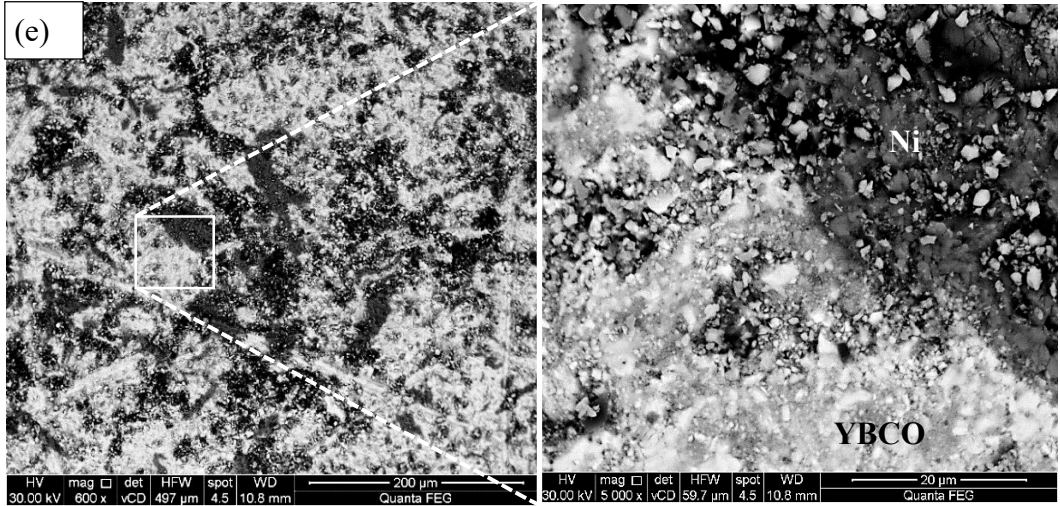
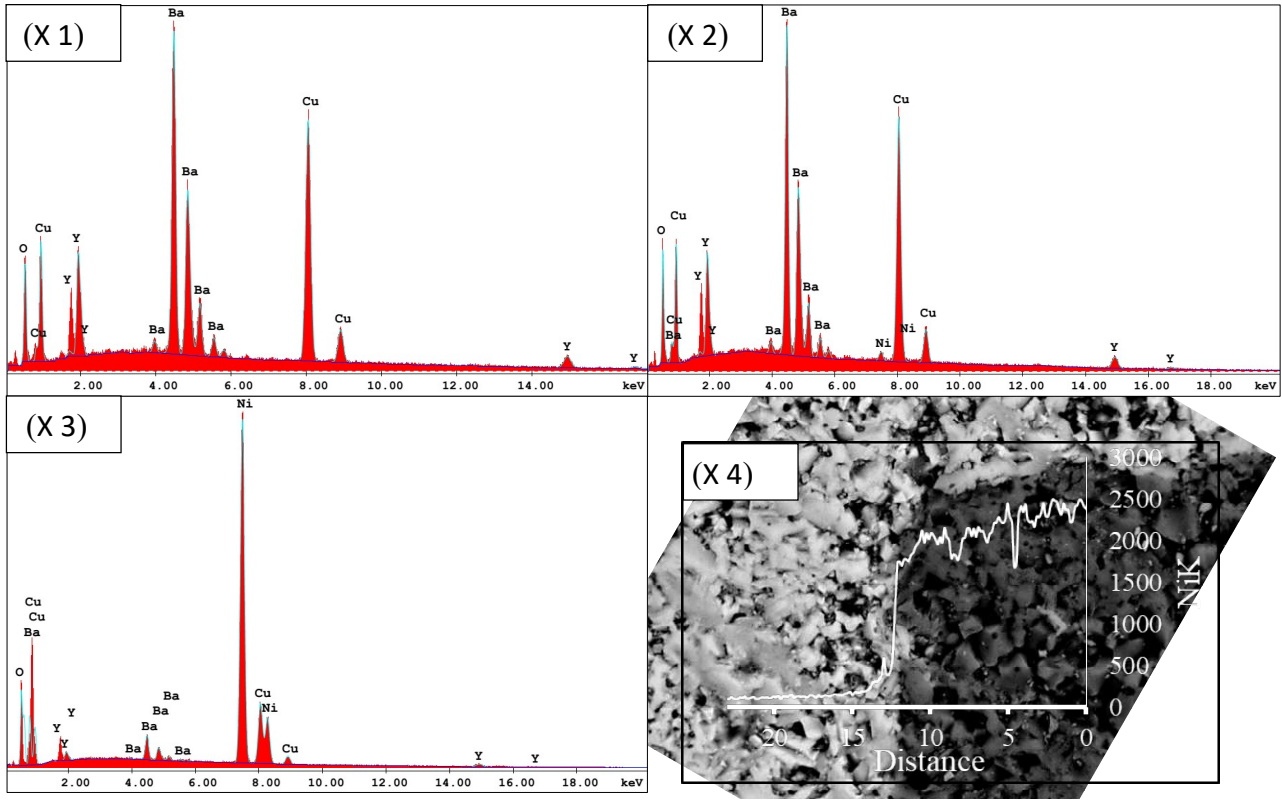

**Figure 3:** SEM micrographs of the samples YBCO + xNi composites (x = 0, 2.5, 5, 10 and 15 wt.%) marked as (a), (b), (c), (d) and (e) ,respectively and EDX graphs of YBCO + 2.5 wt.% Ni samples and linear EDX of 10 wt.% Ni sample at positions marked, X1, X2, X3, X4, respectively.

### 3.3. Temperature dependence of resistivity analysis

Figure 4 shows the measurements of the resistivity, $\rho(\Omega cm)$, and the temperature derivative, $d\rho/dT$ ($\Omega cmK^{-1}$) versus temperature for all the samples. The straight line fits to the normal state resistivity $\rho_n$ are extrapolated from 300K to 0K. From these data, the room-temperature resistivity $\rho_{room}(\Omega cm)$, superconducting critical temperature $T_C$ (K), superconducting transition width $\Delta T_C$ (K) and zero-resistivity-critical temperature ($T_{C0}$) were determined for all the YBCO + xNi composites. For easy comparison, resistivity data of all the samples are graphed together in Figures 4(h) and 4(i). For pure YBCO and the compounds with low Ni content (1 to 5 wt.%), as shown in Figures 4(a-d), the resistive transition exhibits two different regimes. The first is the temperature range from $\sim 2T_C$ to 300 K. The normal state resistivity $\rho_n(T)$ and shows a metallic behavior. It follows a T-linear behavior, $\rho_n(T) = \alpha + \beta T$, over an extended temperature range, characteristic of optimally doped cuprates[37–39]. $\rho_o$ is the residual resistivity which can be obtained from the linear fitting of resistivity in the temperature range from $2T_C$ to 300 K and extrapolating it to 0K, as shown in Figures 4(a-d). The second low temperature region is characterized by the contribution from fluctuating Cooper pairs, where $\rho(T)$ is deviates downward from linearity $\rho_n(T)$[39,40]. This is mainly the paraconductivity dominated regime[39,40]. For samples with high Ni additives (7.5 to 15 wt.%), the resistive transition exhibits semi-metallic and semiconducting behaviors, as shown in Figures 4(e-g).

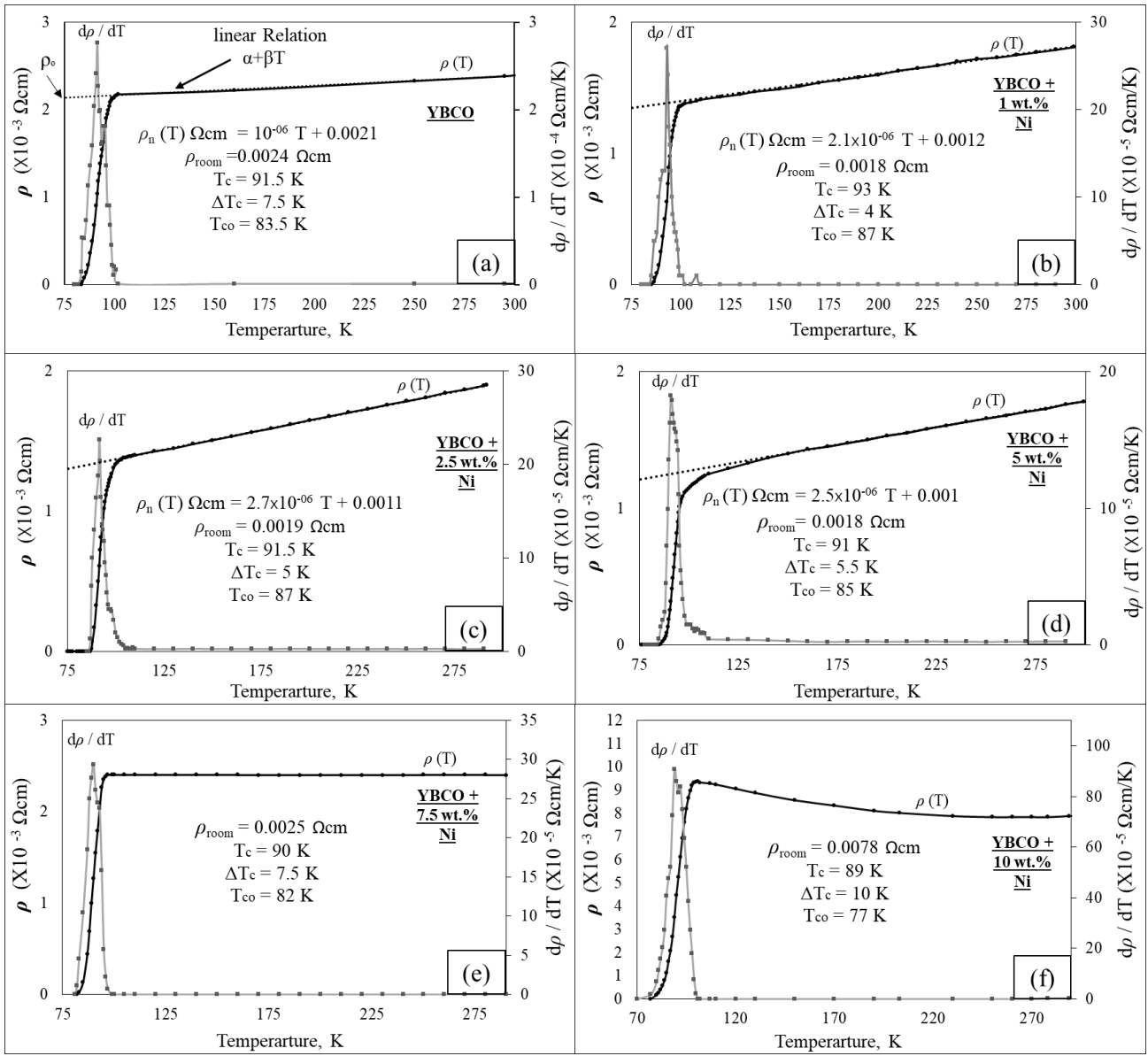

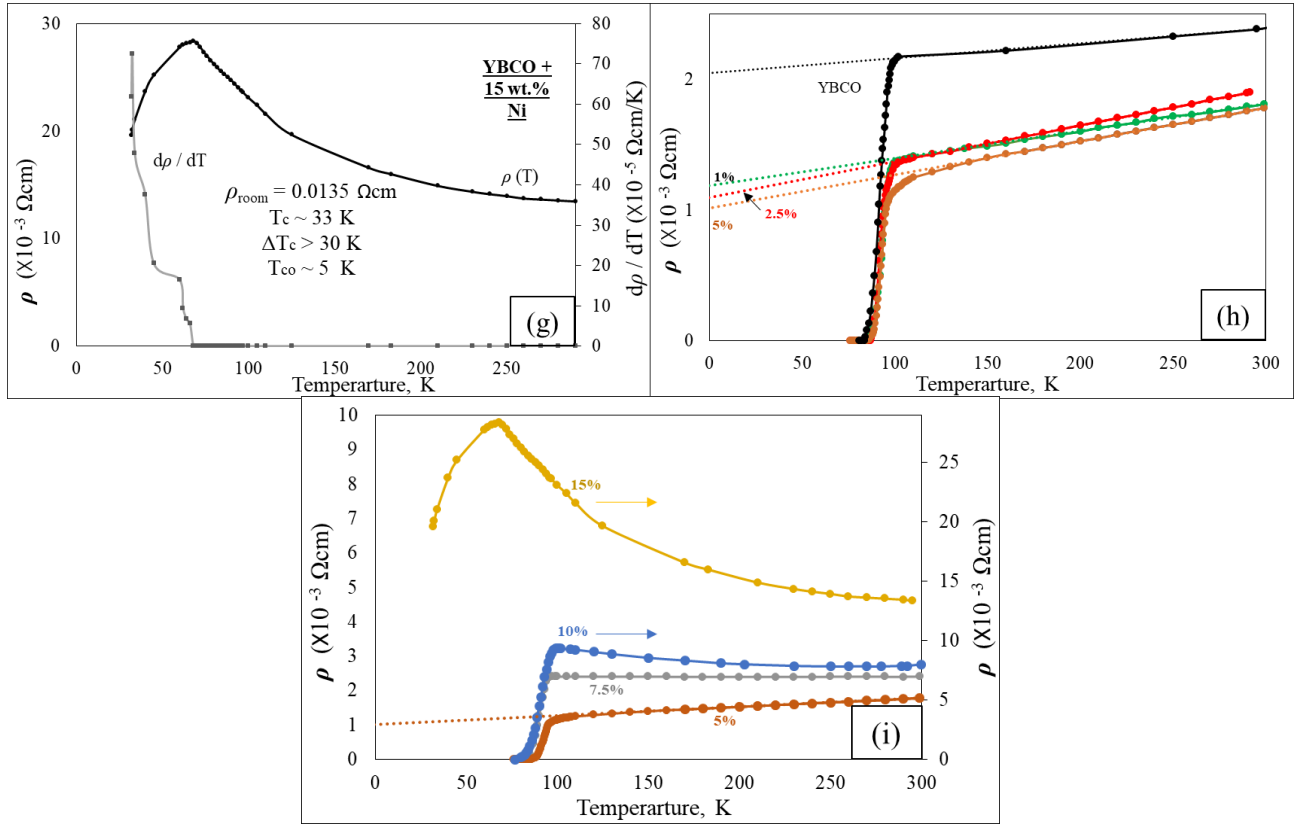

**Figure 4.** The experimentally measured dc resistivity, $\rho$ ($\Omega$cm) and its temperature derivative, $d\rho/dT$ ($\Omega$cmK$^{-1}$) versus temperature. The straight line fit to the normal state resistivity $\rho_n$ extrapolated from room temperature 300K to 0K is also shown. Some relevant superconducting and normal state parameters are given in the plots.

Important normal and superconducting state properties of the composite samples are listed in Table 2. The transition width is determined from the half-width of the $d\rho/dT$ profile. Residual resistivities of all the samples are dominated by very high grain boundary contributions.

**Table 2** Some normal and superconducting state features of YBCO + xNi (x = 0, 1, 2.5, 5, 7.5, 10, 15 wt. %) composites.

| Ni content (wt.%) | $T_C$ (K) | $T_{C0}$ (K) | $\Delta T_C$ (K) | $d\rho/dT$ (10$^{-6}$ $\Omega$cm/K) | $\rho_o$ (10$^{-3}$ $\Omega$cm) | $\rho_{room}$ (10$^{-3}$ $\Omega$cm) |
|---|---|---|---|---|---|---|
| 0 | 91.5 | 83.5 | 7.5 | 1 | 2.1 | 2.4 |
| 1 | 93 | 87 | 4 | 2.1 | 1.2 | 1.8 |
| 2.5 | 91.5 | 87 | 5 | 2.7 | 1.1 | 1.9 |
| 5 | 91 | 85 | 5.5 | 2.5 | 1 | 1.8 |
| 7.5 | 90 | 82 | 7.5 | 0 | 2.5 | 2.5 |
| 10 | 89 | 77 | 10 | - | - | 7.8 |
| 15 | ~ 33 | ~ 5 | > 30 | - | - | 13.5 |

Surprisingly, it is observed that some of the superconducting and normal state properties were improved by increasing Ni content up to 5 wt. %. For example, the metallic behavior, characterized by ($d\rho/dT$), $\rho_{room}$ and $\rho_o$, improved relative to the pure superconducting YBCO, as shown in Figure 4(h) and Table 2. The superconducting $T_C$ also shows improvement in comparison to the pure compound. The sample with 1 wt.% Ni exhibits very high $T_C$ and relatively sharp transition width. At 7.5 wt.%, where c/a ratios are small relative to pure YBCO, there is large potential scattering of hole carriers, and corresponding increasing of resistivity. Thus, the sample behavior starts to change from metallic to semi-metallic and the superconducting properties start to recess, as shown in Figure 4(i) and Table 2. In spite of large c/a ratios at ≥10 wt. %, the increasingly growing NiO phase starts to have a dominant detrimental effect to the overall superconducting properties of the whole system. This behavior is well complemented by the temperature-dependent magnetization data of YBCO + xNi composites.

### 3.4. Magnetization measurements

Figure 5 shows the measurements of magnetic moment (M) as a function of applied magnetic field (H) for the YBCO + xNi (x = 0, 1, 2.5, 5, 10 wt. %) compounds at 79K. The insertion is the maximum moment at applied field ~ | 220 | Gauss vs. Ni wt.%. The magnetization measurement showed the

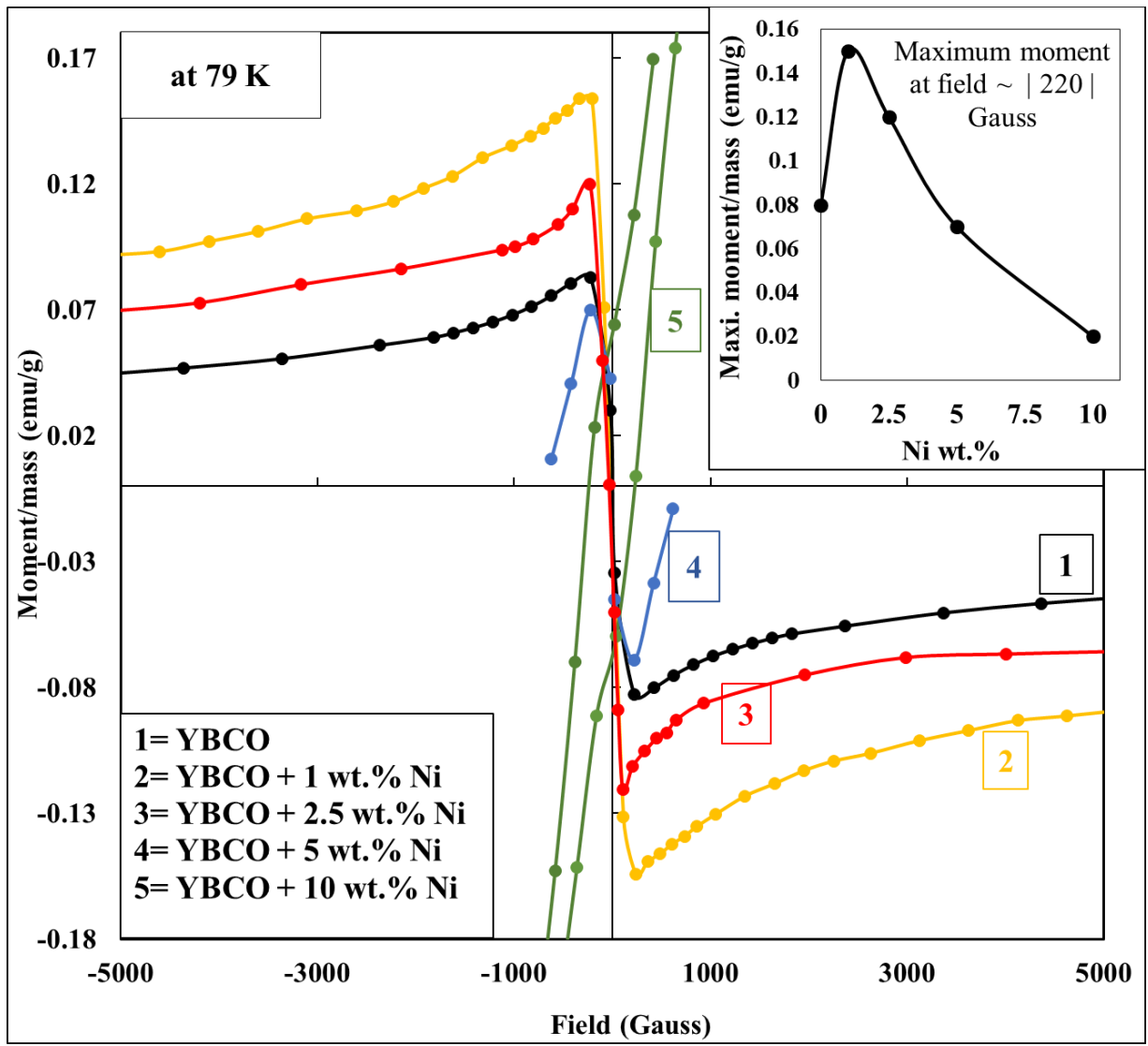

**Figure 5:** Magnetic moment vs. field for the samples YBCO + xNi (x = 0, 1, 2.5, 5, 10 wt.%) at 79K.

increase of the magnetic moment of the YBCO + 1 wt.% Ni composite with subsequent decreasing of the moment of the composite with an increase of Ni wt.% (insertion of Figure 5). The magnetic field lines are being expelled out of the superconducting samples YBCO + xNi (x = 0, 1 and 2.5 wt.%). This signifies that these samples show clear diamagnetic property. Samples with 1 and 2.5 wt.% Ni have much higher M values than that of pure YBCO sample. This is due to the enhanced superconductivity in these two composites. These composites are expected to have significantly higher $J_C$ because NiO additives will act as strong pinning centers without being detrimental to superconductivity itself. YBCO + xNi (x = 5 and 10 wt.%) composites, on the other hand, could initially expel out the flux lines for small magnetic field. Then by increasing the applied field, the system behavior changed from diamagnetic into antiferromagnetic, as a result of the effect of antiferromagnetic NiO inclusions. In other words, increasing Ni gradually reduces the diamagnetic signal and eventually the Meissner effect is completely masked by the antiferromagnetic contributions originating from the large amount of NiO additives. It is interesting to note that $T_C$ values of these two composites are still quite high, ~ 90K.

## 4. Discussion

All the above modifications of the structural and electrical properties of YBCO + xNi composites are direct results of Ni addition in the YBCO matrix (some of the Cu sites in the plane and the chain are also substituted by Ni[31–33,41]. $T_C$ is mainly controlled by the amount of the in-plane $Ni^{2+}$ substitution). $Ni^{2+}$ doping in Cu-plane sites, or in-plane $Ni^{2+}$, can have two different roles; potential and magnetic scatterings.

i) For heavily doped YBCO + xNi, the clear degradations of the superconducting state properties are mainly due to potential scattering[20,42,43]. The in-plane $Ni^{2+}$ modifies the hole hopping integrals $t_{pp}$, $t_{pd}$, $t'_{pp}$ and $t'_{pd}$ [44] as a result of anti-JT distortions (contracted crystallographic c/a ratio) in the YBCO lattice[17]. The contraction of crystallographic c/a ratio alters the degree of hybridization between Cu $3d_{z^2-r^2}$-orbitals and O $2p_z$-orbitals and the energy level of the apical oxygen, $\Delta_A$ gets smaller[44–47]. The apical transfer integrals $t'_{pp}$ and $t'_{pd}$ (or hole hopping in the out-of-plane direction) at $Ni^{2+}$ sites will therefore increase, whereas the planar transfer integrals $t_{pp}$, $t_{pd}$ (or hole hopping in the $CuO_2$ plane) will decrease[47]. In other words, the approach of apical oxygen to $Ni^{2+}$, in the compressed $NiO_5$ pyramids, gives rise to an effect mimicking in-plane pair-breaking by increasing hopping of the holes in the out-of-plane direction. The local distortion around $Ni^{2+}$ is the reason of the impurity scattering potential in this scenario. It's thought that $Ni^{2+}$ ions have no major role on the charge density in the $CuO_2$ plane. As a consequence, the hole density stays nearly unchanged[48,49]. Samples with 1 and 2.5 wt.% Ni are observed to contain YBCO crystallites of increasing c/a ratios (lower $Ni^{2+}$ content) as shown in Figure 2. Increased c/a ratios (implying higher JT distortions in the proposed scenario), result in extended Cu-O apical bonds which in turn increase the energy level of $\Delta_A$. As a result, hopping of holes in the out-of-plane apical direction has smaller values and thus is confined in the $CuO_2$ plane. In other words, the overall superconducting properties of these composites with small amount of Ni additives are improved for dwindling out-of-plane hole hopping. As Ni content increases, unavoidable substitution of $Ni^{2+}$ ions in the Cu-plane sites increase and superconducting state properties are degraded gradually.

ii) The magnetic effect of $Ni^{2+}$ becomes important, especially when the content is high. $Ni^{2+}$ generally has weak effect on antiferromagnetic (AF) superexchange interaction between localized electrons on neighboring Cu-plane sites through intervening $O^{2-}$ ions[50–58]. Zhang and Rice[59] showed that an oxygen $2p_\sigma$-hole with localized spin on Cu-plane site, forms a Kondo-like singlet state[60,61]. This is known as the Zhang–Rice (ZR) singlet, as shown schematically in Figure 6.

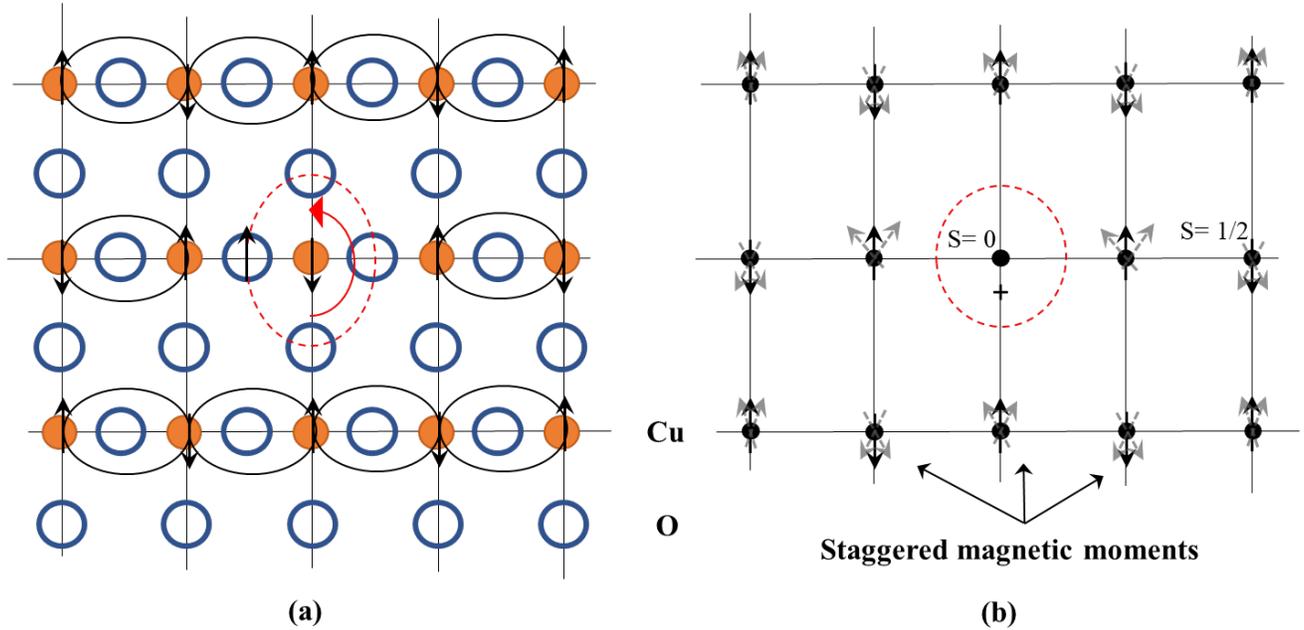

**Figure 6:** (a) Schematic picture of Zhang–Rice singlet. A dopant hole in O $2p_\sigma$ orbitals surrounding a Cu site forms a spin singlet with the localized spin on the Cu site, (b) the formation of a spin singlet with Heisenberg spin system. Note the surrounding staggered magnetic moments due to ZR singlet. Its hopping corresponds to a superexchange process between an itinerant hole and a localized spin.

ZR singlet is a special combination of oxygen $2p_\sigma$-orbitals hybridized with the Cu $3d_{x^2-y^2}$-orbital. ZR singlet is quite stable and influences the AF correlations around it. Particularly, the magnetic moments of the in-plane $Cu^{2+}$ ions neighboring ZR singlet are staggering because of the parallel spins of the neighbor $Cu^{2+}$ ions, see Figure 6(b)[62]. This picture can be replaced by another one as shown in Figure 7(b). The largely staggered magnetic moments at the two neighboring $Cu^{2+}$ ions (of parallel spins) are represented as shown in Figure 7(b).

Pairing of two ZR singlets, as illustrated in Figure 7(c) is relaxing for the lattice to get rid of the perturbation of the staggered magnetic moment. The pairing of ZR singlets comes true due to the coupling (attraction) of the staggered magnetic moments of anti-parallel spins of the two $Cu^{2+}$ ions in singlets close to each other, as shown in Figure 7(c). This pairing process could be disturbed by the lattice vibration. Therefore, by decreasing the temperature towards $T_C$, AF correlated pairing of ZR singlets get stronger. As a result, the system continuously loses its electrical resistance[32,63,64].

The proposed picture of paired ZR singlets could explain effects arising as a result of doping Ni in the $CuO_2$ planes. The electron configuration of $Ni^{+2}$ is $4S^0 3d^8$. It is supposed to have magnetic moment S = 1 (1.73$\mu_B$), but the effective moment is less than this[19,65]. The magnetic moment of $Ni^{+2}$ ion has therefore two components; one is due to $3d^1_{x^2-y^2}$ and the other is to $3d^1_{z^2-r^2}$. The other two nonmagnetic configurations, $3d^2_{x^2-y^2} 3d^0_{z^2-r^2}$ or $3d^0_{x^2-y^2} 3d^2_{z^2-r^2}$, are excluded because of the observed Curie term as reported by Mendels et al[65] for $YBa_2(Cu_{1-y}Ni_y)_3O_{7-\delta}$. $Ni^{2+}$ in the $CuO_2$ plane retains the magnetic moment component of $Ni^{2+}$ $3d^1_{z^2-r^2}$ that will be almost isolated. The magnetic interaction between the $Ni^{2+}$ $3d^1_{z^2-r^2}$ spin and neighboring host $Cu^{2+} 3d^1_{x^2-y^2}$ spins are thought to be very weak.

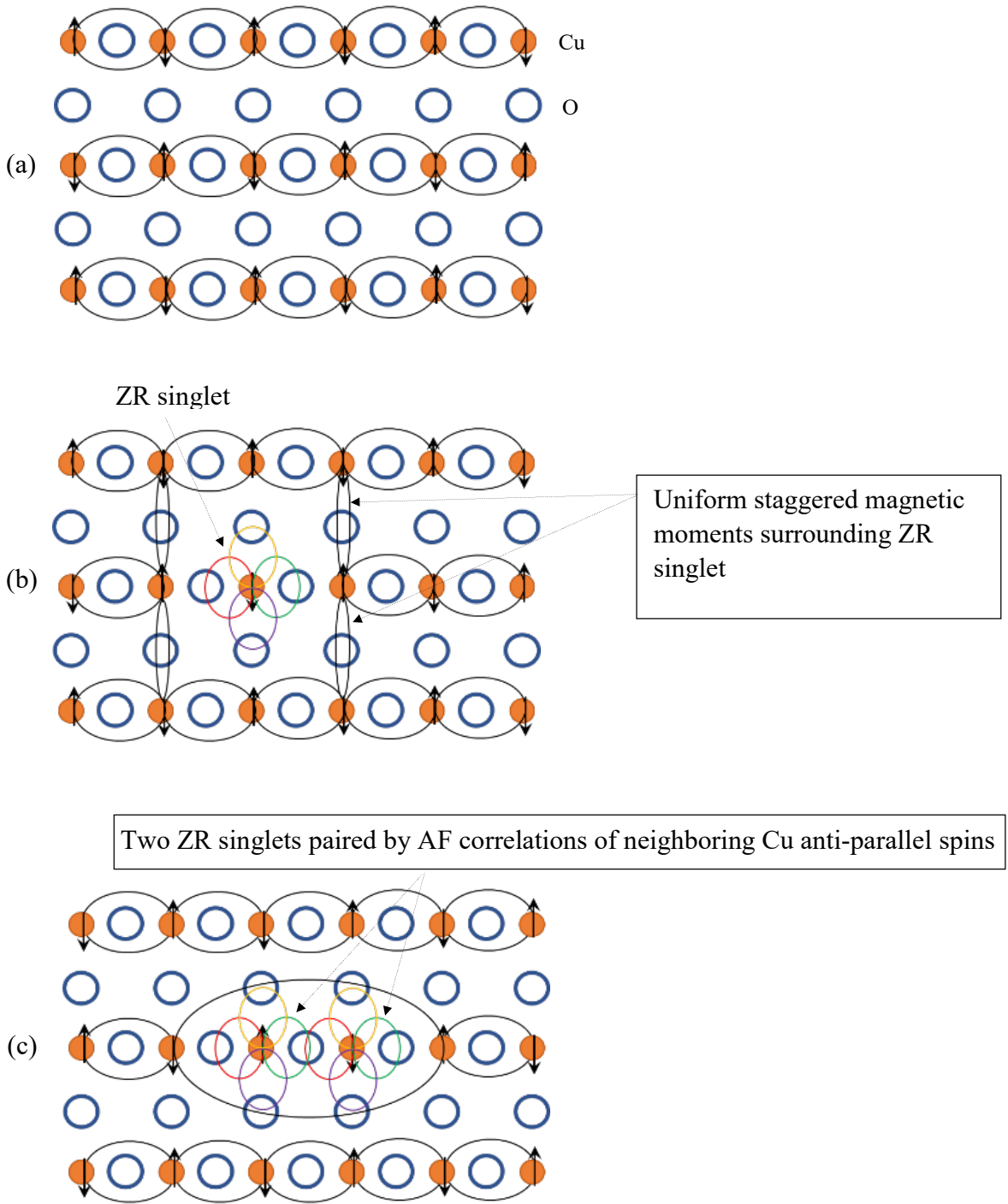

**Figure 7:** (a) Schematic representation of AF superexchange correlation in $CuO_2$ plane of undoped cuprate, (b) the surrounding staggered magnetic moments due to the formation of ZR singlet, colored circles represent possible ZR-singlets associated with itinerancy of doped hole on the four oxygens surrounding a Cu ion with localized spin,(c) the proposed picture of two ZR singlets pairing due to coupling of the surrounding anti parallel spin magnetic moments of the two neighbor Cu ions (the most stable "relaxed" state in a relatively frozen lattice).

Therefore, it is reasonable to assume that isolated spin magnetic moment $Ni^{2+}$ $3d^1_{z^2-r^2}$ barley perturbs the AF correlations, as shown in Figure 8(a). Concerning spin fluctuations, magnetic probe experiments revealed significant changes in the AF correlations with Zn-doping[66–73] but only weak perturbations with Ni-doping[67,70,72,74].

The overall effect of Ni ion in $CuO_2$ plane is intricate. We think that the magnetic moment component of $Ni^{2+}$ $3d^1_{z^2-r^2}$ may have some role in localizing hole carriers in O $2p_z$ orbitals by antiferromagnetic coupling with single holes in these orbitals, but further investigation is needed. Moreover, we expect a scattering process for paired ZR singlets which may produce localized hole states only in the proximity of $Ni^{2+}$ site as shown in Figure 8(b)[22,48,49,73]. In other words, the pairing interactions of ZR singlets are perturbed at the local environment of $Ni^{2+}$ and the subsequent magnetic confinement of hole carriers around $Ni^{2+}$ inhibits Cooper pair formation process. In this situation, relatively high Ni-doping is expected to decrease the concentration of the delocalized hole pairs.

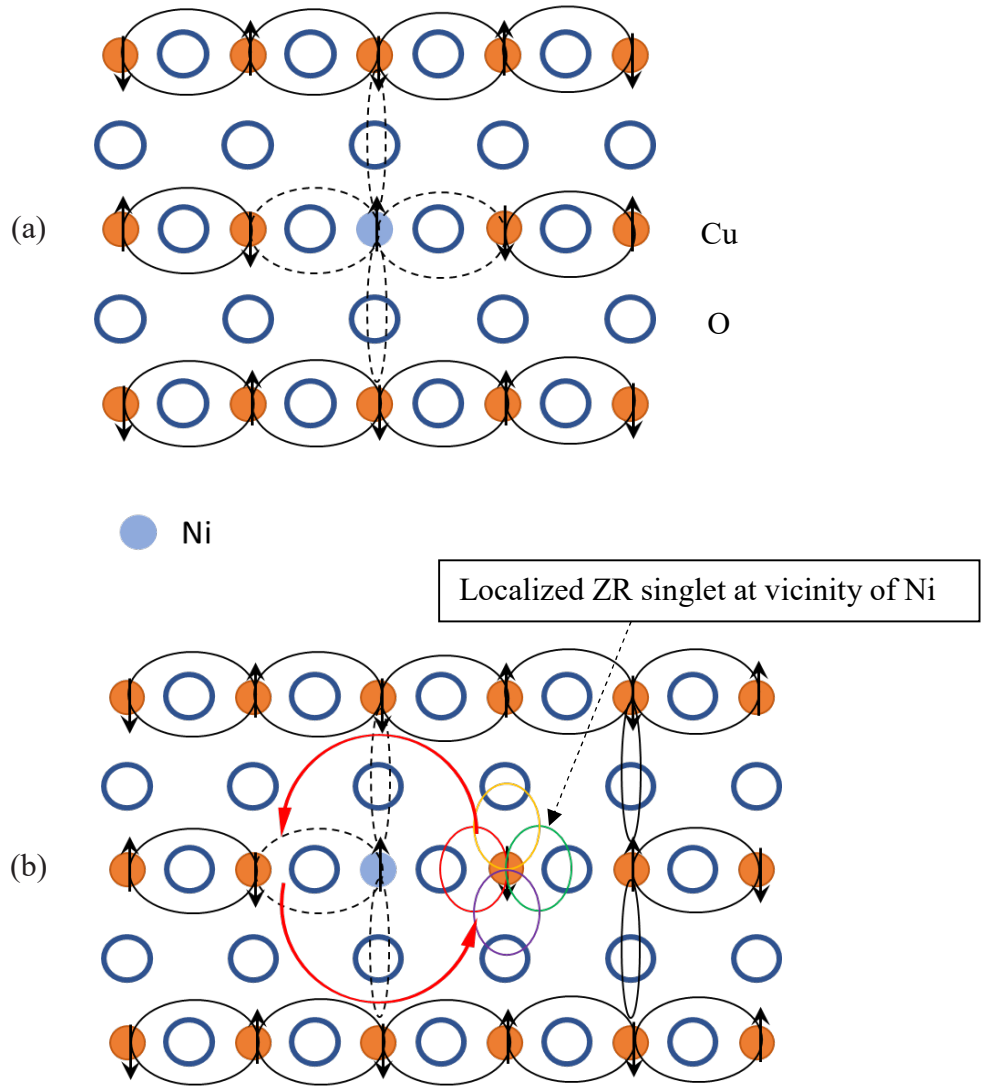

**Figure 8.** (a) Staggered magnetic moment at $Ni^{2+}$ and associated perturbation of surrounding AF superexchange correlation due to isolated spin magnetic moment of Ni $3d^1_{z^2-r^2}$, (b) probable localization of ZR singlet surrounding $Ni^{2+}$ due to the perturbation of AF superexchange correlation.

For low level of Ni addition, namely 1 and 2.5 wt.% Ni (of higher c/a ratios and lower potential scattering), the effect of hole localization is weak. The probability of existence of ZR-singlets pairs in the regions of the $CuO_2$ plane; separated from other $Ni^{2+}$ is high in this situation. As the average separation between $Ni^{2+}$ sites are expected to be much larger than the SC coherence length at low level of Ni addition, $Ni^{2+}$ should be ineffective in carrier localization. By increasing Ni concentration in $CuO_2$ plane, the lattice distortion (and related changes in $t_{pp}$, $t_{pd}$, $t'_{pp}$ and $t'_{pd}$) together with magnetic confinement of doped holes to the $Ni^{2+}$ sites result in strong scattering and hole localization. The role of hole localization process gets more effective at low temperatures. Moreover, the distance between $Ni^{2+}$ sites along $CuO_2$ plane gets smaller as amount of Ni increases. Therefore, the SC wave function starts to scatter as its spatial extent cannot be confined to a region smaller than its coherence length. In other words, it's not possible that the SC state to be constrained to occupy regions of the $CuO_2$ plane in between $Ni^{2+}$ centers. This explains the suppression of the SC properties, namely lower critical magnetic field, $T_C$ and critical current densities and changing from metallic into semi-metallic and then into semiconducting behaviors, despite of increasing c/a ratio again in 7.5, 10 and 15 wt.% YBCO + xNi composites.

## 5. Conclusions

Unexpected improvements of SC state properties of some YBCO + xNi (where x ≤ 2.5 wt.%) has been suggested as due to high JT distortions. These samples are found to have YBCO crystallites with high c/a ratios. We have proposed that these composites have high value of $\Delta_A$ and high in-plane hopping integrals on average. YBCO-crystallites with low level of Ni additives, in-plane $Ni^{2+}$ barely perturbs the short-range AF correlations that are supposed to have effective role in the pairing processes of the ZR singlets. Further addition of $Ni^{2+}$ leads to an unavoidable substitution in the Cu sites which gives rise to localization of in-plane hole carriers in vicinity of $Ni^{2+}$ and in turn scattering of moving ZR singlets-pairs. These result in the degraded state of superconductivity for composites with high level of Ni wt.%.

## Acknowledgements

This work was supported by the Egyptian National Institute of Standards.